\documentclass[prl,superscriptaddress,floatfix,showpacs,aps,reprint,letterpaper,nobalancelastpage]{revtex4-1}
\usepackage{graphicx} 
\usepackage{color} 
\usepackage{transparent}
\usepackage{amsmath}
\usepackage{hyperref} 
\usepackage{amssymb}

\begin{document}

\title{Process verification of two-qubit quantum gates by randomized benchmarking}\date{\today}

\author{A. D. C\'orcoles}
\affiliation{IBM T.J. Watson Research Center, Yorktown Heights, NY 10598, USA}
\author{Jay M. Gambetta}
\affiliation{IBM T.J. Watson Research Center, Yorktown Heights, NY 10598, USA}
\author{Jerry M. Chow}
\affiliation{IBM T.J. Watson Research Center, Yorktown Heights, NY 10598, USA}
\author{John A. Smolin}
\affiliation{IBM T.J. Watson Research Center, Yorktown Heights, NY 10598, USA}
\author{Matthew Ware}
\affiliation{Department of Physics, Syracuse University, Syracuse, NY 13244, USA}
\author{J. D. Strand}
\affiliation{Department of Physics, Syracuse University, Syracuse, NY 13244, USA}
\author{B. L. T. Plourde}
\affiliation{Department of Physics, Syracuse University, Syracuse, NY 13244, USA}
\author{M. Steffen}
\affiliation{IBM T.J. Watson Research Center, Yorktown Heights, NY 10598, USA}

\begin{abstract}
We implement a complete randomized benchmarking protocol on a system of two superconducting qubits. The protocol consists of randomizing over gates in the Clifford group, which experimentally are generated via an improved two-qubit cross-resonance gate implementation and single-qubit unitaries. From this we extract an optimal average error per Clifford of $0.0936$. We also perform an interleaved experiment, alternating our optimal two-qubit gate with random two-qubit Clifford gates, to obtain a two-qubit gate error of $0.0653$. We compare these values with a two-qubit gate error of $\sim0.12$ obtained from quantum process tomography, which is likely limited by state preparation and measurement errors. 
\end{abstract}
\maketitle 

A task of fundamental importance in the development of large-scale quantum processors is that of determining a robust, multi-platform standard to quantify the quality of the quantum operations. Characterizing quantum gates becomes harder as the error of the gates continues to decrease. Quantum process tomography \cite{PCZ,CN}, a standard method for characterizing quantum gates, gives a full description of the protocol under study, but is very sensitive to state preparation and measurement (SPAM) errors, especially when the errors involved are comparable to those of the single qubit unitaries of the system. As an alternative, recent work has been devoted to the development of randomized benchmarking (RB) protocols \cite{EAZ,KLRB,MGE}, which have been implemented in ion traps \cite{KLRB,Biercuk2009,BWC}, NMR \cite{RLL}, superconducting qubits \cite{Chow2010a,CGT,Paik2011} and atoms in optical lattices \cite{Olmschenk2010}. Although a RB protocol offers a reliable estimate of the average error per operator within a group of operators in its original conception, it does not provide a complete description of a given quantum process. However, recent studies \cite{Magesan2012,Gaebler2012} have presented a different RB implementation in which the error of a particular gate is measured by interleaving it between other random gates within a sequence. Furthermore, RB has also been used to determine addressability errors and correlations in many-body quantum systems \cite{Gambetta2012}.

In this Letter, we present a complete RB characterization of two fixed-frequency superconducting qubits. An important feature of a RB protocol is the long sequences required to implement it. This sets additional constraints on the chosen gates, which have to be robust as a part of a long series of pulses in the same way they would in a real quantum algorithm. In our case, the realization of the RB protocol is made possible by modifying a previously developed cross-resonance (CR) two-qubit gate \cite{Chow2011,Chow2012} through refocusing the single-qubit dynamics to simplify the gate calibration.

Our implementation of the RB protocol is restricted to the Clifford group $\mathcal{C}$ of operators, which is the normalizer of the Pauli group $\mathcal{P}$ -that is, for each $C \in \mathcal{C}$, $CPC^{\dagger} \in \mathcal{P}$, with $P \in \mathcal{P}$. Having the Clifford group as our set of operators is justified by two main reasons. First, a Clifford-based RB protocol is more readily extendable to systems with higher number of qubits, as choosing a random Clifford and decomposing it into a set of generators (elementary qubit operations) scales polynomially in the number of qubits \cite{MGE,AG}. Second, as the set of generators tends to vary across systems, such a protocol offers portability between the different physical implementations of quantum processors. 

For two-qubit systems, the Clifford group is generated from single-qubit unitaries and a controlled-NOT (CNOT) gate. When implementing the protocol, it is important to use a Clifford decomposition into elementary unitaries that minimizes the number of average two-qubit gates per Clifford, as these tend to have lower fidelities than single-qubit gates, especially in the case of superconducting qubits. Rather than following the procedure outlined in Ref. \cite{MGE} for generating a random Clifford operation, here we use an optimized set of Clifford operations from which we randomly select elements. We divided the two-qubit Clifford group into four classes \cite{sup}: a class with 576 elements containing only single-qubit unitaries from the group $\{I,X_{\pm\pi/2},Y_{\pm\pi/2},X_{\pi},Y_{\pi}\}$, where $U_{\theta}$ represents a rotation of angle $\theta$ around the axis $U$, a class with 5184 elements containing single-qubit unitaries and one CNOT gate, a class with 5184 elements containing single-qubit unitaries and two CNOT gates and a class with 576 elements containing single-qubit unitaries and a SWAP gate, implemented by three CNOT gates. Therefore, the total number of elements in the two-qubit Clifford group is $11520$, with an average of $1.5$ CNOT gates per Clifford \cite{cliff}.
\begin{figure}[t!]
\centering
\includegraphics[width=0.48\textwidth]{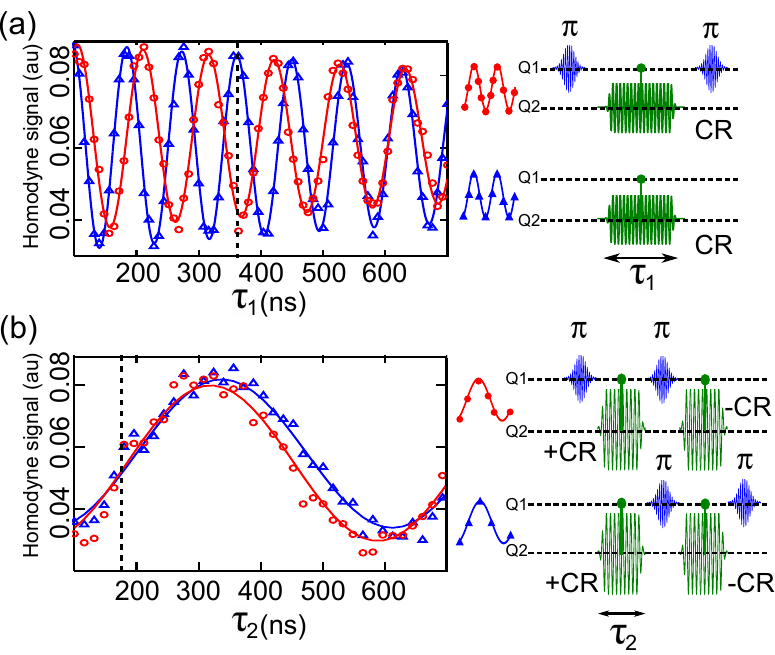}
\caption{\label{fig1}(color online) Homodyne signal as a function of the cross-resonance tone length and pulse sequences for the standard cross-resonance gate (a) and for the $ZX_{-\pi/2}$ (b). All tones are applied to the control qubit driving line (Q1), with the $\pi$ pulses and cross-resonance tones applied at the control and target qubit frequencies, respectively. The $\pi$ pulse between the two cross-resonance tones in the $ZX_{-\pi/2}$ gate, along with the opposite sign of the second cross-resonance pulse, echo away the fast $IX$ component of the driving Hamiltonian seen in (a) and leave only the $ZX$ interaction. The dashed vertical lines mark the gate length used for the RB experiments for the $ZX_{-\pi/2}$ gate and the equivalent length for the standard cross-resonance gate.}

\end{figure}

The experiments are performed on two single-junction transmons coupled via a superconducting coplanar waveguide resonator, which is also used for readout \cite{Blais:2004}. The qubit resonance frequencies are $\omega_1/2\pi = 3.2324$ and $\omega_2/2\pi = 3.2945$ GHz, with anharmonicities $\delta_1=(\omega_{1}^{12}-\omega_{1}^{01})/2\pi=-331$ and $\delta_2=(\omega_{2}^{12}-\omega_{2}^{01})/2\pi=-216$ MHz, whereas the bare resonator frequency is $\omega_{\rm{r}}/2\pi = 8.2855$ GHz. The energy relaxation times of both qubits are observed to be $T_1^{(1)}=11.6$ $\mu$s and $T_1^{(2)}=9.1$ $\mu$s, and the coherence times observed from Ramsey-fringe experiments are $T_2^{(1)}=7.1$ $\mu$s and $T_2^{(2)}=5.6$ $\mu$s \cite{sup}. The qubits are thermally anchored to the coldest stage of a dilution refrigerator with a nominal base temperature of 15 mK and are carefully shielded against thermal radiation \cite{Barends2011,Corcoles2011}. 

Implementing the Clifford group of operators in our experiments required modifying a previously demonstrated two-qubit CR gate \cite{Chow2011}. The basis of the original CR gate \cite{Paraoanu2006,Rigetti2010} involves the driving of the control qubit at the frequency of the target qubit. This results in a driving Hamiltonian of the form 
\begin{equation}
\label{Eq1}
H_D/\hbar \approx \epsilon(t)\Big( mIX -\mu ZX + \eta ZI\Big)
\end{equation}
where $\{I,X,Y,Z\}^{\otimes2}$ are the two-qubit Pauli operators, $\epsilon$ is the drive amplitude, $\mu$ is a coupling parameter that equals $J/\Delta$ for ideal qubits, where $J$ is the qubit-qubit coupling energy and $\Delta$ is the frequency detuning between the qubits, $m$ is a scalar representing the effect of spurious electromagnetic crosstalk between both qubits as well as the effect of higher energy levels, and $\eta$ represents the magnitude of the Stark shift arising from the off-resonant driving of qubit 1. The term $m IX$ results in Rabi-like oscillations of qubit 2, to which the term $-\mu ZX$ contributes with a slower rotation whose sign depends on the state of qubit 1. The effect of the Hamiltonian $H_D$ can be seen in the experiments shown in Fig. \ref{fig1} (a). A CR flattop Gaussian pulse of variable length $\tau_1$ and Gaussian width $\sigma=8$ ns is applied to qubit 1 (control) at the frequency of qubit 2 (target). Depending on whether a $\pi$ rotation is applied to qubit 1 prior to the CR pulse (circles) or not (triangles), different Rabi rates are observed on qubit 2. In the experiments where qubit 1 is in the excited state prior to the CR pulse, an additional $\pi$ rotation is applied to qubit 1 at the end of the sequence in order to have the Rabi-oscillation signal from the joint readout between the same two computational states $|00\rangle$ and $|01\rangle$. All single-qubit rotations are Gaussian-shaped pulses with a total length of 32 ns and $\sigma=8$ ns. To avoid leakage into higher energy levels, all single-qubit pulses include a calibrated derivative in the other quadrature \cite{Motzoi:2009,Chow2010a}.

The Hamiltonian in Equation \ref{Eq1}, however, presents difficulties for implementing long sequences of Cliffords due to the single-qubit terms. These terms could be explicitly tuned out with additional simultaneous pulses, but this would be rather demanding on the phase-locking and amplitude stability requirements of the electronics and on sequence complexity which would have an important negative impact on the measured gate fidelity. Instead, we construct a more manageable two-qubit Clifford by modifying the original pulse sequence in order to remove all terms except the one corresponding to $ZX$. The new pulse sequence divides the CR pulse in three parts: an initial CR pulse of duration $\tau_2$, a $\pi$ rotation of qubit 1, and a final CR pulse of opposite sign to the first one, also of duration $\tau_2$. As in the original pulse scheme, a final $\pi$ rotation is applied to qubit 1 when pertinent. The net effect of the new pulse sequence is to effectively ``echo'' away the terms involving $IX$ and $ZI$ in $H_D$ and, as a result, only the slower Rabi-rotation arising from $ZX$ is observed [Fig. \ref{fig1} (b)]. In the experiments presented here, the gate is realized by chosing a value of $\tau_2$ that leaves qubit 2 in a superposition of $|0\rangle$ and $|1\rangle$. We will therefore call this gate $ZX_{-\pi/2}$, the minus sign arising from the negative term involving $ZX$ in Equation \ref{Eq1}. The dashed line in Fig. \ref{fig1} (b) at $\tau_2=178$ ns shows the duration of each of the CR pulses for the $ZX_{-\pi/2}$ gate for these data. The total $ZX_{\pi/2}$ gate length is, therefore, $2\times\tau_2 + 2\times32 = 420$ ns. The equivalent gate length for the original CR scheme is likewise shown in Fig. \ref{fig1} (a) at $\tau_1=356$ ns.

\begin{figure}[t!]
\centering
\includegraphics[width=0.48\textwidth]{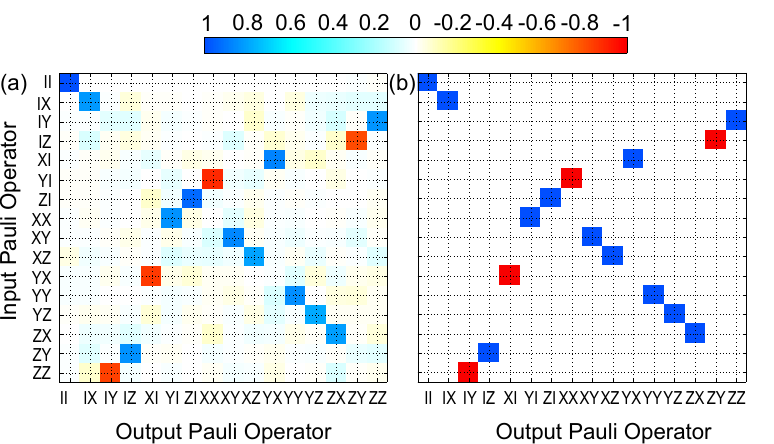}
\caption{\label{fig2}(color online) Quantum process tomography of the $ZX_{-\pi/2}$ gate with $\tau_2 = 178$ ns. (a) Experimentally extracted Pauli transfer matrix. The gate fidelity is $F_g = 0.8830$ raw and $F_{\rm{mle}}=0.8799$ after applying a maximum likelihood algorithm. (b) Ideal Pauli transfer matrix representation of the $ZX_{-\pi/2}$ gate.}

\end{figure}

We performed quantum process tomography on the 178 ns $ZX_{-\pi/2}$ gate (Fig. \ref{fig2}) by preparing an overcomplete set of 36 states generated by $\{I,X_{\pi},X_{\pm\pi/2},Y_{\pm\pi/2}\}$, applying the gate to each of them and performing state tomography. We use the Pauli basis to represent quantum process tomography through the Pauli transfer matrix $\mathcal{R}$ \cite{Chow2012}. The raw data is post-processed with a semidefinite algorithm \cite{Chow2012} to take into account physicality constraints such as complete positivity and trace preservation of the process. We obtain a gate fidelity from the raw data of $F_g=0.8830$ and a maximum-likelihood estimated fidelity of $F_{\rm{mle}}=0.8799$. These results, as the RB measurements described next suggest, are probably dominated by SPAM errors.

We base our implementation of a two-qubit RB protocol on the theory described in Refs. \cite{MGE} and \cite{Gaebler2012}. We randomly choose a sequence $\{C_1, C_2,...,C_{20}\}$ of 20 Cliffords amongst the 11520 elements of the two-qubit Clifford group. From this sequence, we then construct a series of truncations $\{m_1,m_2,...,m_{20}\}$, where $m_i=\{C_1,C_2,...,C_i\}$. In order to make each truncation self-inverting, a final pulse $I_i$ is appended at the end of each of them which returns the system to the $|00\rangle$ state. The state of the two qubits is then read out. A whole RB experiment, therefore, consists of a series of pulse trains $\{N_1,N_2,...,N_{20}\}$, where $N_i = \{m_i,I_i\}$.

\begin{figure}[t!]
\centering
\includegraphics[width=0.48\textwidth]{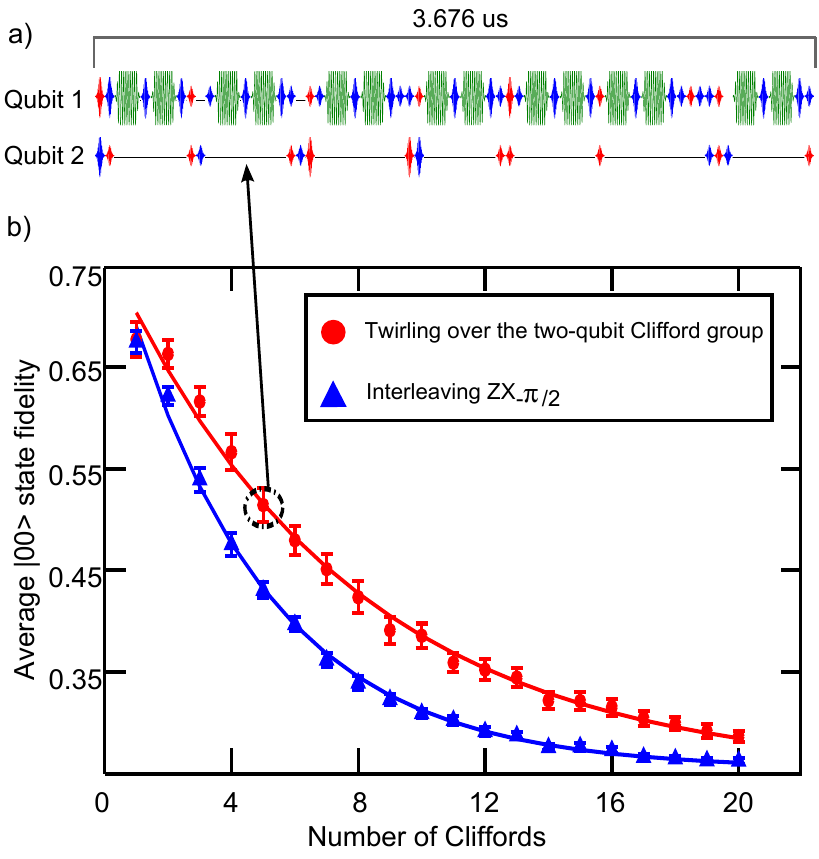}
\caption{\label{fig3}(color online) RB experiments on a two-qubit system. (a) Two-qubit pulse sequence for 5 Cliffords randomly selected from the 11520 elements of the two-qubit Clifford group. Tall Gaussians represent $\pi$ rotations, whereas short ones represent $\pi/2$ rotations. A final Clifford is added to make the sequence self-inverting. (b) Fidelity decay for the $|00\rangle$ in the standard RB protocol where the twirling is performed over the two-qubit Clifford group (circles) and in the interleaved protocol (triangles), in which each Clifford has an additional $ZX_{-\pi/2}$ gate appended at the end. The decays are fitted to an exponential model to extract an average error per Clifford of $r=0.0936\pm0.0058$ (standard protocol) and a $ZX_{-\pi/2}$ error of $r_{\mathcal{C}}=0.0653\pm0.0014$. The arrow shows the truncation that would correspond to the pulse sequence shown in (a) and which has an average $|00\rangle$ state fidelity of over 50\%.}

\end{figure}

The fidelity of the $|00\rangle$ state after each train of pulses $N_i$, obtained by joint readout of the two qubits \cite{Majer2007}, can be fit to an exponential model $F(i,|00\rangle)=A\alpha^i+B$, and the average error rate per Clifford is related to $\alpha$ by the relation $r=1-\alpha-(1-\alpha)/d$, where $d=2^n$ for $n$ qubits \cite{MGE}. In this model, state preparation and measurement errors are absorbed by the constants $A$ and $B$ and therefore the parameter $\alpha$ provides a SPAM-free estimation of the average error per Clifford. One of the shortcomings of this model as described, however, is that it only gives an estimation of the average error per Clifford generator.

We can, nonetheless, modify the protocol in order to estimate the error of a particular gate of interest, $\overline{C}$, also belonging to the Clifford group, as described in Refs. \cite{Magesan2012} and \cite{Gaebler2012}. In the new protocol, we interleave random elements of the Clifford group between the gate under study $\overline{C}$. This is achieved by constructing similar random sequences of Cliffords as in the original implementation and then appending the gate $\overline{C}$ after each element in the sequence. A final inverting Clifford is added at the end of each truncation of the sequences.

Fig. \ref{fig3} (b) shows the average $|00\rangle$ state fidelity over random sequences of Cliffords with lengths ranging from 1 to 20 for both the standard (circles) and the interleaved (triangles) RB protocols. Each data point was averaged over 40 different random sequences. Both sets of RB data in Fig. \ref{fig3} (b) are well described by an exponential model, with reduced $\chi^2$ values of 0.589 and 0.386 for the standard and interleaved RB protocols, respectively. Our construction of the Clifford group results in 30 two-qubit gates on average for a 20 Clifford long sequence. The pulse sequence of one particular self-inverting 5-Clifford long sequence, comprising 7 two-qubit and 23 single-qubit gates, is shown in Fig. \ref{fig3} (a). In the case of the standard RB implentation experiment, we obtain $\alpha=0.8752\pm0.0078$, which results in an average error per Clifford of $r=0.0936\pm0.0058$. For the interleaved experiment, $\alpha_{\mathcal{C}}=0.7990\pm0.0058$, from which $r_{\mathcal{C}}$ can be estimated as $r_{\mathcal{C}}=(d-1)(1-\alpha_{\mathcal{C}}/\alpha)/d  = 0.0653\pm0.0014$ \cite{Magesan2012}. All errors here represent a $1\sigma$ confidence interval obtained from the Jacobian of the fitting model \cite{Gambetta2012}. The average error per Clifford for the single-qubit Clifford group for each qubit was also measured and observed to be $r_1=0.0041\pm0.0001$ for qubit 1 and $r_2=0.0048\pm0.0002$ for qubit 2 \cite{sup}.

The effective coupling strength of the two qubits, given by the product $\epsilon(t) \mu$ multiplying the $ZX$ interaction in Eq. \ref{Eq1}, can be increased with the amplitude of the CR driving tone $\epsilon(t)$. Thus, larger driving amplitudes result in a faster evolution of the system and, therefore, faster oscillations in Fig. \ref{fig1} (b) and shorter $\tau_2$. The ultimate limit for the speed of the $ZX_{-\pi/2}$ gate is determined by the frequency of the oscillations induced by the $ZX$ term, the qubit-qubit detuning $\Delta$, and the qubit anharmonicities. For the strongest drives, energy leakage into other levels in the system prevent the gate from becoming faster \cite{Chow2011,Groot2012}. 

\begin{figure}[t!]
\centering
\includegraphics[width=0.48\textwidth]{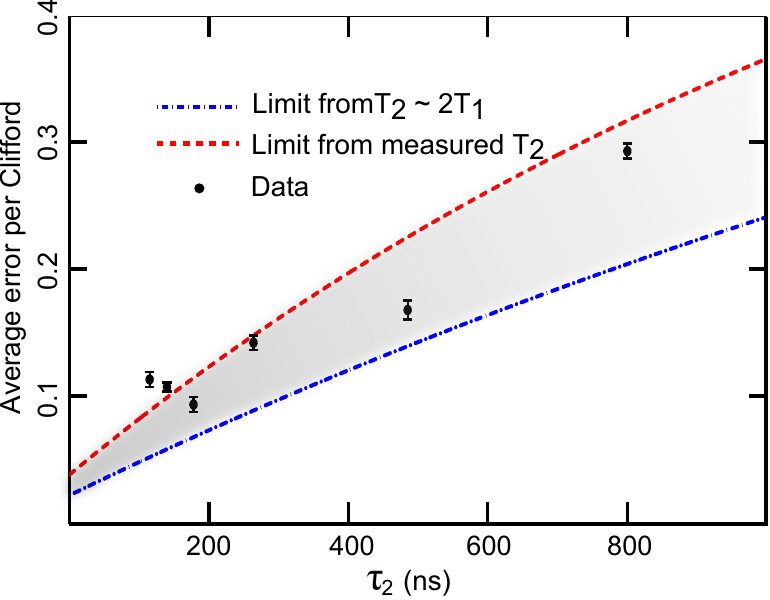}
\caption{\label{fig4}(color online) Average error per Clifford as a function of the $ZX_{-\pi/2}$ gate length $\tau_2$. the shaded area delimits coherence-limited errors and is defined by the calculated error for the measured coherence times of both qubits (dashed line) and the calculated error for a coherence time equal to twice the qubits relaxation times (dash-dotted). Error bars represent $1\sigma$ confidence intervals. These results suggest that our average gate error is currently limited by qubit coherence.}

\end{figure}

We applied the two-qubit RB protocol to our system for different $ZX_{-\pi/2}$ lengths, with $\tau_2$ ranging from 115 to 800 ns. Fig. \ref{fig4} shows the average error per Clifford as a function of $\tau_2$ and the calculated coherence-limited average error for the measured $T_2$ of both qubits (dashed line) and for $T_2^{(1,2)}=2T_1^{(1,2)}$ (dash-dotted line). We see that the experimentally extracted average errors fall inside the shaded region. For the two shortest realizations of the $ZX_{-\pi/2}$, however, the RB experiments yielded an average error per Clifford worse than the limit imposed by coherence times. This was probably due to leakage into higher qubit levels at high CR driving amplitudes and to the spurious off-resonance driving of the control qubit. For longer gates, the observed experimental values indicate that our $ZX_{-\pi/2}$ is essentially coherence limited. We attribute the scattering of the data points in Fig. \ref{fig4} to small variations in $T_1$ and $T_2$, which were observed to move by about 1 $\mu$s up or down for both qubits over the interval of six to twelve hours, approximately equal to the amount of time taken to perform the RB protocol for each $ZX_{-\pi/2}$ gate length.

In conclusion, we have implemented a RB protocol on two superconducting qubits by using a new implementation of an entangling two-qubit gate plus single-qubit unitaries. The new gate pulse sequence echoes away the terms in the gate Hamiltonian involving single-qubit effects, which tend to be non-reproducible in long gate sequences. The new gate implementation allows the experimental realization of the two-qubit RB protocol for the first time in superconducting qubits. Furthermore, an interleaved RB experiment in which the two-qubit gate is alternated with a random Clifford during the sequences yielded a gate fidelity of $0.9347$, which compares favorably to the fidelity of $0.8799$ obtained by quantum process tomography performed on the same gate. Our results show the importance of having a protocol insensitive to errors arising from SPAM, especially as the processes to benchmark become higher and higher in fidelity. Measurements of our protocol as a function of the gate duration suggest that our gate is currently limited by the coherence time of our qubits.

\begin{acknowledgments}
We acknowledge discussions and contributions from Blake Johnson, Colm Ryan, Seth Merkel, Erik Lucero and Douglas McClure. We acknowledge use of the Cornell NanoScale Facility, a member of the National Nanotechnology Infrastructure Network, which is supported by the National Science Foundation (Grant ECS-0335765). We acknowledge support from IARPA under contract W911NF-10-1-0324. All statements of fact, opinion or conclusions contained herein are those of the authors and should not be construed as representing the official views or policies of the U.S. Government.
\end{acknowledgments}

\end{document}


\title{Supplementary material for `Process verification of two-qubit quantum gates by randomized benchmarking'}\date{\today}

\author{A. D. C\'orcoles}
\affiliation{IBM T.J. Watson Research Center, Yorktown Heights, NY 10598, USA}
\author{Jay M. Gambetta}
\affiliation{IBM T.J. Watson Research Center, Yorktown Heights, NY 10598, USA}
\author{Jerry M. Chow}
\affiliation{IBM T.J. Watson Research Center, Yorktown Heights, NY 10598, USA}
\author{John A. Smolin}
\affiliation{IBM T.J. Watson Research Center, Yorktown Heights, NY 10598, USA}
\author{Matthew Ware}
\affiliation{Department of Physics, Syracuse University, Syracuse, NY 13244, USA}
\author{J. D. Strand}
\affiliation{Department of Physics, Syracuse University, Syracuse, NY 13244, USA}
\author{B. L. T. Plourde}
\affiliation{Department of Physics, Syracuse University, Syracuse, NY 13244, USA}
\author{M. Steffen}
\affiliation{IBM T.J. Watson Research Center, Yorktown Heights, NY 10598, USA}

\maketitle 
\section{Randomized benchmarking of single qubit gates}
\begin{figure}[h!]
\centering
\includegraphics[width=\textwidth]{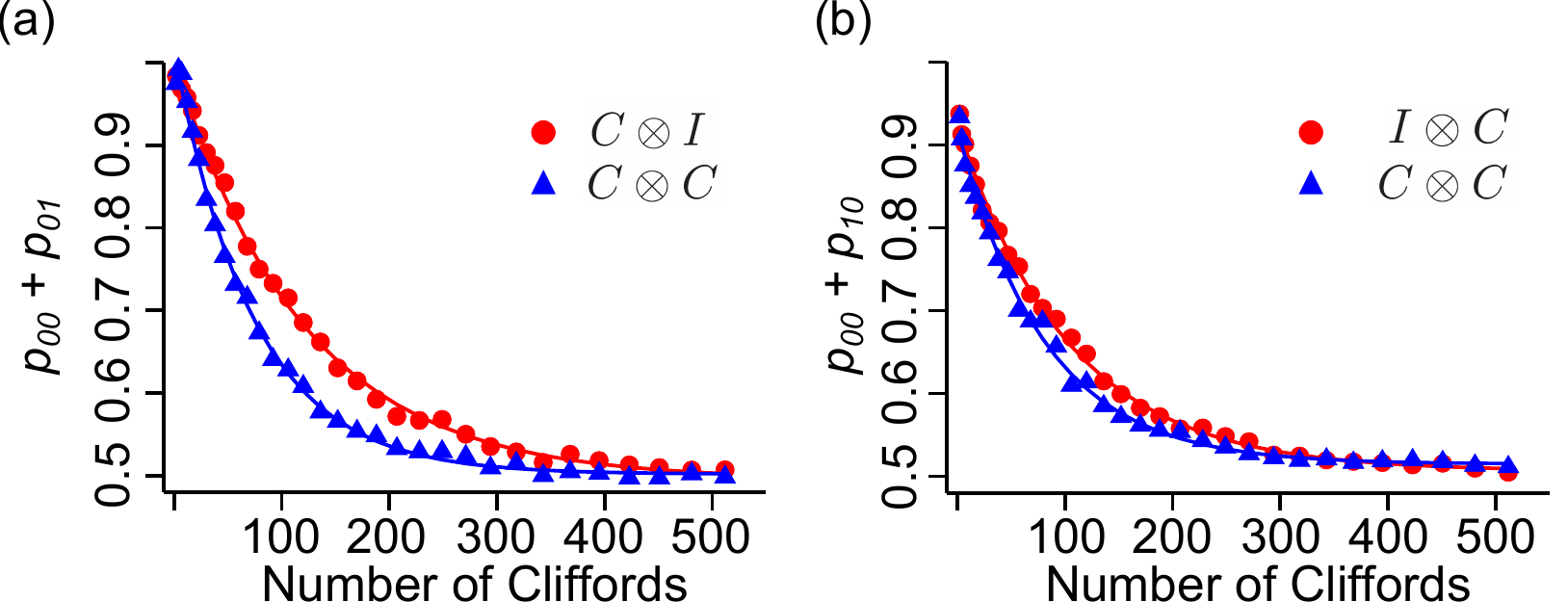}
\caption{\label{fig1sup}(color online) Experimental results of the single-qubit RB experiments. Qubit 1 (a) shows an average error per Clifford of $r_1=0.0041\pm0.0001$ when measured individually (circles) and $r_{1|2}=0.0067\pm0.0002$ when measured simultaneously with qubit 2 (triangles). Qubit 2 (b) has an average error per Clifford of $r_2=0.0048\pm0.0002$ individual (circles) and $r_{2|1}=0.0062\pm0.0002$ (triangles) simultaneous. Uncertainties represent $1\sigma$ confidence intervals.}

\end{figure}

We show here an implementation for our two-qubit system of the RB protocol for single qubit gates described in Ref. \cite{Gambetta2012}. The measurements (see Fig. \ref{fig1sup}) consist of standard RB experiments on each qubit separately ($C\otimes I$ for qubit 1, $I\otimes C$ for qubit 2) and on both qubits simultaneously ($C\otimes C$). The former experiments gives the average error per single-qubit Clifford for each of the qubits, whereas the latter contains information about the amount of spurious crosstalk present in the system. The data in Fig. \ref{fig1sup} were averaged over 20 random sequences for each sequence length.

The decay of each single subsystem is fitted to the same exponential model as the two-qubit RB experiments, $F(i)=A\alpha^i+B$, where $i$ is the number of Cliffords. The results are summarized on Table~\ref{table:1}, where $\alpha_i$ is extracted from the RB experiments individually performed on qubit $i$, and $\delta\alpha=\alpha_{12}-\alpha_{1|2}\alpha_{2|1}$, with $\alpha_{12}$, $\alpha_{1|2}$ and $\alpha_{2|1}$ obtained from fitting $p_{00}+p_{11}$, $p_{00}+p_{01}$ and $p_{00}+p_{10}$, respectively, in the simultaneous RB experiments.
\begin{table}[h!]
\begin{ruledtabular}
\begin{tabular}{cc|cc}
& Twirl Group & Extracted metrics & Reduced $\chi^2$ \\
\hline
$\alpha_1$ & $C\otimes I$ & $0.9918\pm 0.0002$ & 1.371 \\
$\alpha_2$ & $I\otimes C$ & $0.9904 \pm 0.0003 $  & 0.377\\
$\alpha_{1|2}$ & $C\otimes C$ & $0.9865 \pm 0.0003$ & 0.339\\
$\alpha_{2|1}$ & $C\otimes C$ & $0.9876\pm 0.0004$  & 0.243\\
$\alpha_{12}$ & $C\otimes C$ & $0.9745 \pm 0.0011$ & 0.705\\
$\delta \alpha$ & - & $0.0002 \pm 0.0018 $  & -\\
\end{tabular}
\end{ruledtabular}
 \caption{\label{table:1} Summary of the extracted data from the single-qubit RB experiments on both qubits. Uncertainties represent $1\sigma$ confidence intervals.}
\end{table}

The average errors per Clifford $r_i=(1-\alpha_i)/2$ obtained from the above results are $r_1=0.0041\pm0.0001$ and $r_{1|2}=0.0067\pm0.0002$ for qubit 1 and $r_2=0.0048\pm0.0002$ and $r_{2|1}=0.0062\pm0.0002$ for qubit 2. The parameter $\delta \alpha$ represents the amount of correlations in the system, a measure of entangling errors. From our experiments we can see that, although there is some amount of crosstalk ($\delta r_{1|2} = r_{1|2}-r_1 = 0.0026\pm0.0003$ and  $\delta r_{2|1} = r_{2|1}-r_2 = 0.0014\pm0.0004$) in our sample, the amount of entangling error is minimal. This reflects favourably on the performance of the $ZX_{\pi/2}$ gate, which turns out to be coherence limited and not affected by systematic errors.

\section{Qubit parameters}

The relaxation and coherence times of both qubits are quoted as the average of 120 independent measurements. For qubit 1, the average of the relaxation time measurements was $T_1^{(1)}=11.6$ $\mu$s with standard deviation $\sigma_{T1}^{(1)}=1.6$ $\mu$s and the average of the coherence time measurements was $T_2^{(1)}=7.1$ $\mu$s with standard deviation $\sigma_{T2}^{(1)}=4.7$ $\mu$s For qubit 2, $T_1^{(2)}=9.1$ $\mu$s with standard deviation $\sigma_{T1}^{(2)}=0.9$ $\mu$s and $T_2^{(2)}=5.6$ $\mu$s with standard deviation $\sigma_{T2}^{(2)}=0.5$ $\mu$s.

The qubits anharmonicities were $(\omega_{1}^{12}-\omega_{1}^{01})/2\pi=-331$ MHz for qubit 1 and $(\omega_{2}^{12}-\omega_{2}^{01})/2\pi=-216$ MHz for qubit 2.

\section{Decomposition of the two-qubit Clifford Operations}

Defining  $ \mathcal{C}_1$ as the group of single qubit Clifford operators (which has 24 different elements), the two-qubit Clifford group can be found with the help of the group $\mathcal{S}_1=\{I, R_S,R_S^2\}$ where $R_S$ is the Pauli transfer matrix of the unitary operator $S=\exp[-i (X+Y+Z)\pi/\sqrt{3}3]$. This group is simply the rotation that exchanges all the axes of the Bloch sphere (the $x-y-z$ axis maps to $y-z-x$). To do this we note that there are four distinct classes of the two-qubit Clifford group. The first class consists of 576 elements ($24^2$) and represents all single qubit Clifford operations
\[ \Qcircuit @C=1em @R=.7em {
&\gate {\mathcal{C}_1}&\qw\\
& \gate{\mathcal{C}_1}&\qw
}\]
The second class has 5184 elements  ($24^2\times 3^2$)  and contains all combinations of the following sequence
\[ \Qcircuit @C=1em @R=.7em {
& \gate{\mathcal{C}_1}& \ctrl{1} & \gate{\mathcal{S}_1} &\qw\\
& \gate{\mathcal{C}_1}& \targ & \gate{\mathcal{S}_1} &\qw
}\] We call this the CNOT-like class. The third class also has 5184 elements ($24^2\times 3^2$) and contains all combinations of the following sequence 
\[ \Qcircuit @C=1em @R=.7em {
& \gate{\mathcal{C}_1}& \iswap  & \gate{\mathcal{S}_1} &\qw\\
& \gate{\mathcal{C}_1}&  \iswap\qwx  &\gate{\mathcal{S}_1} &\qw
}\] We call this the iSWAP-like class. It should be noted that we are using a non-standard notation for the iSWAP gate.  The final class is the SWAP class and consists of all 576 ($24^2$) combinations of the following sequence 
\[ \Qcircuit @C=1em @R=.7em {
& \gate{\mathcal{C}_1}& \qswap &\qw\\
& \gate{\mathcal{C}_1}& \qswap \qwx&\qw
}\] This is the optimal decomposition of the two-qubit Clifford group in terms of number of CNOTs as it can be shown that implementing a iSWAP requires two CNOTs and a SWAP requires three. Thus on average the number of  CNOTs to implement a two-qubit Clifford is 1.5. The same is also true if the building block was the iSWAP as it takes two of these to make a CNOT and three to make a SWAP \cite{Schuch2003}. The number of single qubit gates depends on how the single qubit Cliffords are implemented. It can not be less than an average of 3.8 single qubit gates for each two-qubit Clifford and in general it will be more. This is because, for our system, it is simpler to tune up a finite set of generators and decompose all Clifford operations in terms of them. For this experiment the generating set was  $\{I, X_{\pm\pi/2}, Y_{\pm\pi/2}, X_\pi, Y_\pi\}^{\otimes 2}$, and $ZX_{-\pi/2}$. Since we can simply incorporate any single qubit operations into the pre $\mathcal{C}_1$ and any element form $\mathcal{S}_1$ into the post rotation we use the following replacements for the two-qubit entangling gates  of the above classes 
\[ \Qcircuit @C=1em @R=.7em {
&\ctrl{1} &\qw & \push{\rule{.3em}{0em}\rightarrow\rule{.3em}{0em}} && \multigate{1}{ZX_{-\pi/2}} &\qw \\
&\targ &\qw & &&  \ghost{ZX_{-\pi/2}} & \qw
}\]
\[ \Qcircuit @C=1em @R=.7em {
&\iswap&\qw & \push{\rule{.3em}{0em}\rightarrow\rule{.3em}{0em}} && \multigate{1}{ZX_{-\pi/2}}&  \gate{Y_{-\pi/2}}& \multigate{1}{ZX_{-\pi/2}} &\qw \\
&\iswap\qwx&\qw & &&  \ghost{ZX_{-\pi/2}}& \gate{Y_{-\pi/2}}&  \ghost{ZX_{-\pi/2}} & \qw
}\]
and for the SWAP class we use the replacement

\[ \Qcircuit @C=1em @R=.7em {
&\qswap &\qw & \push{\rule{.3em}{0em}\rightarrow\rule{.3em}{0em}} && \multigate{1}{ZX_{-\pi/2}}&  \gate{Y_{-\pi/2}}& \multigate{1}{ZX_{-\pi/2}} &\gate{X_{\pi/2}}&\qw& \multigate{1}{ZX_{-\pi/2}}&\qw \\
&\qswap\qwx&\qw & &&  \ghost{ZX_{-\pi/2}}& \gate{Y_{-\pi/2}}&  \ghost{ZX_{-\pi/2}} &\gate{X_{\pi/2}}&\gate{Y_{-\pi/2}}& \ghost{ZX_{-\pi/2}} &\qw
}\]

With these decompositions we find the average number of $ZX_{-\pi/2}$ is 1.5 and the number of single qubit gates is either 5.6 or 7 depending on whether the identity operations are included as a single qubit rotation.  This is basicially the same as that used in Ref. \cite{Gaebler2012}.